\documentclass{emulateapj}
\newcommand{\gps}{\ensuremath{g_{\rm P1}}}

\newcommand{\PS}{\protect \hbox {Pan-STARRS1}}
\newcommand{\degree}{\ensuremath{^\circ}}
\newcommand{\dfplot}[1]{\plotone{#1}}

\newcommand{\T}{\ensuremath{\vec{\theta}}}
\newcommand{\m}{\ensuremath{\vec{m}}}
\renewcommand{\mp}{\ensuremath{\left\{ \m \right\}}}
\renewcommand{\a}{\ensuremath{\vec{\alpha}}}
\newcommand{\mum}{\ensuremath{\mu}m}

\shorttitle{Dust Reddening to 4.5~kpc}
\shortauthors{E. F. Schlafly et al.}
\begin{document}
\title{A Map of Dust Reddening to 4.5~kpc from Pan-STARRS1}
%
% PS1 paper authorship lists major paper contributors, followed by alphabetical list of PS1 builders,
% You'll want to shuffle the affiliation designations as needed. PS1 institutional addresses are 
% provided below. Look here to get the most up-to-date list of builders and their institutional affiliations:
% http://ps1sc.ifa.hawaii.edu/PS1wiki/index.php/PS1_Builders_aastex
%
% Note that any authors not from PS-1 institutions,  that is not a PS1 builder, needs to secure external
% scientist status. Information on that process is available at
% http://ps1sc.ifa.hawaii.edu/PS1wiki/index.php/Pending_External_Scientists
%
% This example has a first author from UH:
\author{
E. F. Schlafly,\altaffilmark{1}
G. Green,\altaffilmark{2}
D. P. Finkbeiner,\altaffilmark{2,3}
M. Juri\'c,\altaffilmark{4}
H.-W. Rix,\altaffilmark{1}
N. F. Martin,\altaffilmark{5,1}
W. S. Burgett,\altaffilmark{6}
K. C. Chambers,\altaffilmark{6}
P. W. Draper,\altaffilmark{7}
K. W. Hodapp,\altaffilmark{6}
N. Kaiser,\altaffilmark{6}
R.-P. Kudritzki,\altaffilmark{6}
E. A. Magnier,\altaffilmark{6}
N. Metcalfe,\altaffilmark{7}
J. S. Morgan,\altaffilmark{6}
P. A. Price,\altaffilmark{8}
C. W. Stubbs,\altaffilmark{3}
J. L. Tonry,\altaffilmark{6}
R. J. Wainscoat,\altaffilmark{6}
C. Waters,\altaffilmark{6}
% this bracket terminates author list
}

% The ordering here should be sequential, matching the sequence in the list of authors:
\altaffiltext{1}{Max Planck Institute for Astronomy, K\"{o}nigstuhl 17, D-69117 Heidelberg, Germany}
\altaffiltext{2}{Harvard-Smithsonian Center for Astrophysics, 60 Garden Street, Cambridge, MA 02138}
\altaffiltext{3}{Department of Physics, Harvard University, 17 Oxford Street, Cambridge MA 02138}
\altaffiltext{4}{LSST Corporation, 933 North Cherry Avenue, Tucson, AZ 85721}
\altaffiltext{5}{Observatoire Astronomique de Strasbourg, CNRS, UMR 7550, 11 rue de l'Universit\'{e}, F-67000 Strasbourg, France}
\altaffiltext{6}{Institute for Astronomy, University of Hawaii, 2680 Woodlawn Drive, Honolulu HI 96822}
\altaffiltext{7}{Department of Physics, Durham University, South Road, Durham DH1 3LE, UK} 
\altaffiltext{8}{Department of Astrophysical Sciences, Princeton University, Princeton, NJ 08544, USA} 
\begin{abstract}
We present a map of the dust reddening to 4.5~kpc derived from Pan-STARRS1 stellar photometry.  The map covers almost the entire sky north of declination $-30\degree$ at a resolution of $7^\prime$--$14^\prime$, and is based on the estimated distances and reddenings to more than 500 million stars.  The technique is designed to map dust in the Galactic plane, where many other techniques are stymied by the presence of multiple dust clouds at different distances along each line of sight.  This reddening-based dust map agrees closely with the \citet[SFD]{Schlegel:1998} far-infrared emission-based dust map away from the Galactic plane, and the most prominent differences between the two maps stem from known limitations of SFD in the plane.  We also compare the map with {\it Planck}, finding likewise good agreement in general at high latitudes.  The use of optical data from Pan-STARRS1 yields reddening uncertainty as low as 25 mmag $E(B-V)$.
\end{abstract}


%% \vfil
%% \eject
%% \clearpage

\section{Introduction}
\label{sec:intro}

Dust absorbs and scatters ultraviolet through infrared light, and reemits the absorbed energy thermally in the mid- through far-infrared.  These processes reshape the radiation field of the Galaxy, playing a crucial role in many areas of Galactic astrophysics  and additionally obscuring and masking astronomical sources observationally.

Because of this influence, maps of dust and its properties are widely used in astronomy.  Dust maps based on neutral hydrogen gas \citep{Burstein:1978} and, more recently, on thermal dust emission \citep[SFD]{Schlegel:1998} have enjoyed wide use.  The advent of wide field digital surveys like the Sloan Digital Sky Survey \citep[SDSS]{York:2000} and the 2 Micron All Sky Survey \citep[2MASS]{Skrutskie:2006} have allowed dust reddening, as opposed to emission, to be mapped accurately on large scales for the first time.  Recent SDSS papers on the topic include works using the photometry of stars \citep{Schlafly:2010, Berry:2012} and galaxies \citep{Yasuda:2007, Peek:2010} as well as spectroscopy \citep{Schlafly:2011, Yuan:2013}.  Studies of dust based on data from 2MASS include photometric studies of reddening \citep{Lombardi:2001, Rowles:2009, Majewski:2011, Nidever:2012} as well as others using number counts \citep{Dobashi:2005} or combinations of both \citep{Marshall:2006}.  The work of \citet{Lallement:2014} maps the three-dimensional structure of the dust within about 1~kpc.  These works have provided new constraints on the nature of the dust reddening law, the variation in dust properties over the sky, and the distribution of mass in clouds, to mention a few results.

The advent of {\it Planck} data makes this an especially interesting time for dust mapping \citep{Planck:2011b, Planck:2013}.  The {\it Planck} data provide full-sky coverage of dust emission from 353--857 GHz.  Large-scale reddening maps complement this data, allowing the dust to be studied simultaneously in both emission and extinction.

As part of a larger effort to study the three-dimensional distribution of dust using PS1 photometry, we present here a new map of dust reddening using photometry of stars observed by \PS\ \citep{PS1_system}.  The map is constructed using the technique of \citet{Green:2014}, in which the distance and reddening to each star is inferred based on its observed photometry, assuming a standard Galactic extinction law with $R_V = 3.1$ \citep{Fitzpatrick:1999}.  Along each $7^\prime\times7^\prime$ line of sight ($14^\prime$ when $|b| > 30\degree$), we model the reddening as a function of distance to infer the reddening to 4.5~kpc, where the reddening is particularly well constrained.  This map has an unprecedented combination of sky coverage, sensitivity, and resolution compared to earlier direct measurements of dust reddening, and is available at our web site\footnote{http://faun.rc.fas.harvard.edu/eschlafly/2dmap}.  The work of \citet{Schlafly:2014} has shown the accuracy of the distances estimated by our approach; this work serves to demonstrate the performance of our reddening estimation.

We describe in \textsection\ref{sec:data} the \PS\ survey, which provides the observation on which our map is based.  We then present our technique and the resulting reddening map in \textsection\ref{sec:method} and \textsection\ref{sec:map}.  We compare it the \citet{Schlegel:1998} dust map and the {\it Planck} dust map in \textsection\ref{sec:discussion}.  We conclude in \textsection\ref{sec:conclusion}.

\section{The \PS\ Survey}
\label{sec:data}

The \PS\ survey, now reaching completion, has observed the entire sky north of declination $-30\degree$ in five filters covering 400--1000~nm \citep{PS_lasercal, JTphoto}.  The 1.8~meter PS1 telescope has a 7 square degree field of view outfitted with a billion-pixel camera \citep{PS1_optics, PS1_GPCA, PS1_GPCB}, and reaches a $5\sigma$ single epoch depth of about 22.0, 22.0, 21.9, 21.0 and 19.8 magnitudes in $grizy_{\rm P1}$. The survey pipeline automatically processes images and performs photometry and astrometry on detected sources \citep{PS1_IPP, PS1_photometry, PS1_astrometry}.  The photometric calibration of the survey is better than 1\% \citep{Schlafly:2012}.

We infer the dust reddening to 4.5~kpc from estimates of the distance and reddening to all stars well-observed by \PS.  For our purposes, we define ``well-observed'' to mean that the star has been observed in good conditions in at least the \gps\ filter and three of the other four PS1 filters.  We use average PS1 photometry in each band; we ignore the variability information encoded in the multiple PS1 observations of each object.  We use point source photometry, and reject galaxies by demanding that the aperture magnitudes of the objects be brighter than the psf magnitudes by less than 0.1~mags in at least three bands, a criterion chosen to yield a clean stellar locus at high Galactic latitudes.  The resulting catalog contains more than 500 million stars over three-quarters of the sky.

\begin{figure*}[htb]
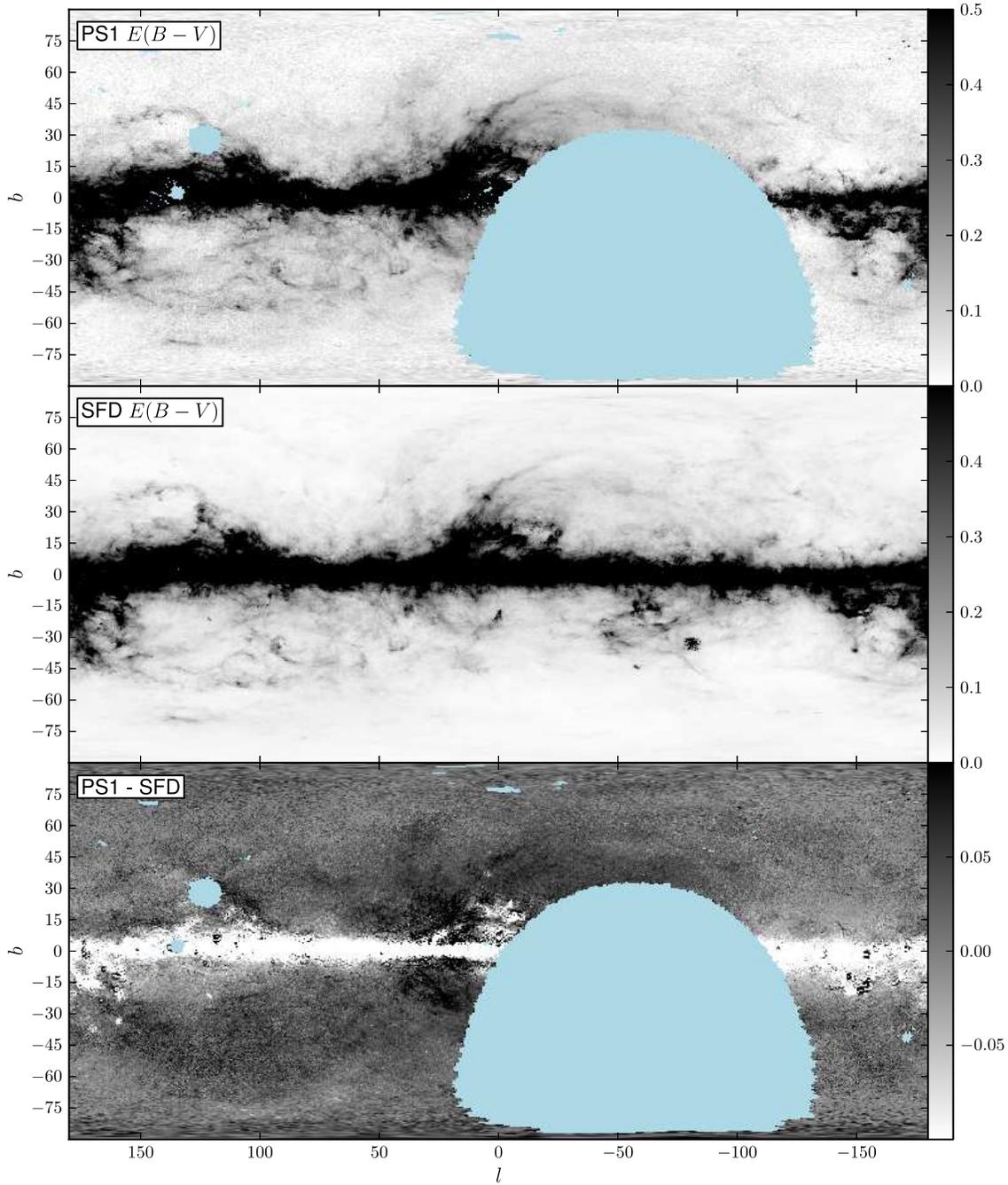

\dfplot{ext-sfd-3panel}
\figcaption{
\label{fig:map3panel}
The maximum-likelihood PS1-based map of the dust at 4.5~kpc; the SFD dust map, which estimates the total reddening from its far-infrared emission; and the difference.  Outside the Galactic plane, the maps agree closely.  Light blue areas denote regions not yet well-observed by PS1; the large blank region is the declination $-30\degree$ boundary.  The full resolution map is available at our web site.
}
\end{figure*}

\section{Method}
\label{sec:method}

The observed photometry of a star depends on the light emitted by the star itself, and also on the distance and reddening to the star.  Accordingly, we model the photometry of each star individually, with parameters for the distance, reddening, and type of the star, as described in \citet{Green:2014}.  We assume here a fixed $R_V=3.1$ reddening law from \citet{Fitzpatrick:1999}, translated into the PS1 bands by \citet{Schlafly:2011}.  We use an empirical set of models for the PS1 colors using fits to the shape of the PS1 stellar locus from \citet{Green:2014}.  These colors are associated with absolute magnitudes and metallicities according to the work of \citet{Ivezic:2008}.  We determine the full probability distribution function for each star's distance, reddening, absolute magnitude, and metallicity.  In this analysis, we impose priors on the distribution of stars and their types using the Galactic model of \citet{Juric:2008} and the metallicity study of \citet{Ivezic:2008}.  We adopt a luminosity function assuming a \citet{Chabrier:2001} initial mass function and using the PARSEC stellar evolution models \citep{Bressan:2012, Green:2014}.  By marginalizing out the absolute magnitude and metallicity, we obtain the posterior probability distribution function $p(E,D|\m)$ describing the range of possible reddenings $E$ and distances $D$ to each star with PS1 photometry $\m$.  We note that all these priors are spatially smooth and described by only around twenty parameters, and carry no information about the detailed distribution of the dust.  This technique is similar to those of \citet{Berry:2012}, \citet{Sale:2009}, \citet{Sale:2012}, \citet{BailerJones:2011} and \citet{Hanson:2014}.

We use a HEALPix $N_{\mathrm{side}} = 512$ pixelization of the sky \citep{Gorski:1999} to split the sky into equal-area $7^\prime \times 7^\prime$ lines of sight ($14^\prime \times 14^\prime$ for $|b| > 30\degree$, $N_{\mathrm{side}} = 256$), and consider the stars along each such line of sight together.  We find the reddening profile $E(D)$ that is most consistent with these stars.  We parameterize $E(D)$ as an increasing, positive, piecewise-linear function in distance modulus $\mu = 5 \log (D / 10~\mathrm{pc})$, with parameters $\a$ giving the reddening at each anchor point of $E(D)$.  The anchors are evenly spaced in $\mu$, with 30 anchors from $\mu = 4.22$ to $\mu = 19.68$.  We then find the maximum likelihood $\a$ given the observed photometry $\mp$ on each line of sight; that is, we maximize $p(\a\ | \mp)$.  As shown in detail by \citet{Green:2014} and summarized here, this simplifies to the product over integrals through $p(E,D|\m)$ for each star $i$, because
\begin{equation}
\label{eq:bestfit}
p(\a | \mp) \propto p(\a) \prod_i p(\m_i | \a) \\
\end{equation}
where we have used Bayes' rule and assumed the photometry of each star along the line of sight to be independent.  Meanwhile
\begin{eqnarray}
p(\m | \a) & = & \int dD d\T p(\m, D, \T | \a) \\
& = & \int dD d\T p(\m | D, \T, \a) p(D, \T) \\
& = & \int dD d\T p(\m | D, \T, E(D ; \a)) p(D, \T)
\end{eqnarray}
where \T\ gives the intrinsic parameters describing each star and $E(D ; \a)$ gives the reddening profile described by parameters $\a$.  The function $p(D, \T)$ incorporates our prior knowledge about the luminosity function and the spatial distribution of stars and their metallicities.  The integral $\int d\T p(\m | D, \T, E) p(D, \T)$ is, up to a normalizing constant, the same as $p(E, D | \m)$ when a flat prior on $E$ is adopted.  This makes $p(\a | \mp)$ ultimately the product of line integrals over $p(E, D | \m)$.

The resulting $p(\a | \mp)$ on each line of sight describe the full three-dimensional distribution of the dust, though in the radial direction the resolution is much worse than in the angular direction.  Additionally the radial direction is more vulnerable to systematic errors from mismatch between our model photometry and the observed photometry.  In this work we present an estimate of the cumulative reddening out to $D = 4.5$~kpc from our three-dimensional map.  This corresponds to $E(4.5~\mathrm{kpc}; \a)$ for the parameters $\a$ that maximize $p(\a)$ along each line of sight.  The reddening to this distance is particularly well constrained by the PS1 photometry, as it corresponds approximately to the single-epoch $g$-band PS1 completeness limit through 1 mag $E(B-V)$ for main-sequence turn-off stars.  We estimate the uncertainty in $E(4.5~\mathrm{kpc})$ by finding the range of allowed reddening to 4.5~kpc such that $\Delta \log p > -0.5$ (i.e., $\Delta \chi^2 < 1$), while holding the reddening at other distances fixed insofar as possible, subject to the constraint that the reddening must increase with distance.  We defer the analysis of the full 3D distribution of the dust to later work.

At Galactic latitudes with $|b| > 30\degree$, occasionally the fitting process becomes unreliable due to the small number of available stars in each pixel.  Accordingly, for the $|b| > 30\degree$ sky, we use a lower $N_{\mathrm{side}} = 256$ pixelization, with $14^\prime$ pixels.  This ensures reliable fits over the entire $\delta > -30$ sky observed by \PS.

\section{Reddening Map}
\label{sec:map}

We present in Figure~\ref{fig:map3panel} a map of the reddening to 4.5~kpc of the entire sky north of declination $-30\degree$, derived from \PS\ photometry.  Also shown is the \citet[SFD]{Schlegel:1998} dust map, which is based entirely on the thermal emission from dust in the far-infrared (FIR), and the difference between the two maps.

Away from the midplane of the inner Galaxy, the two maps agree well.  The large discrepancy in the inner Galaxy is expected because the PS1-based dust map measures the reddening out to only 4.5~kpc, while the SFD map includes the entire reddening through the Galaxy.  More than five degrees off the Galactic plane, where most dust is within about 1~kpc, both maps essentially estimate the full reddening and agree.  This result is expected, but remarkable: though SFD is based on satellite observations of the thermal emission of dust in the FIR, and the PS1 map measures the reddening of optical starlight passing through the dust, the two maps are extremely similar.

The map has $7^\prime$ resolution for $|b| < 30\degree$ and covers about three quarters of the sky.  The noise in the map depends on the local stellar density, the PS1 depth in the region, and the reddening profile.  In the best cases, in regions of low reddening but large stellar density, our map and SFD agree with an rms scatter of only 25 mmag per $7^\prime$ pixel (see \textsection\ref{subsec:sfd}).

Our map has a typical formal uncertainty of about $\sim 20$ mmag $E(B-V)$.  However, the PS1 map uncertainty estimates depend significantly on how aggressively we reduce the weight of outliers in the analysis.  To make a more empirical internal uncertainty estimate, we can compare the reddening map at 4.5~kpc with the reddening map at 2.8 and 7.4~kpc.  When $|b| > 30\degree$, we expect almost all the dust to be within 1~kpc, and the difference of the 2.8 and 7.4~kpc maps is then a sign of the uncertainty in the map. These two maps have a mean difference of about 30 mmag $E(B-V)$, with a scatter similarly of 30 mmag.  We therefore budget a systematic error estimate of 30 mmag.  This is not far from our formal error estimates, but is more realistic, particularly in light of our comparisons with emission-based maps in \textsection\ref{subsec:sfd}.  This estimate is based on the high-latitude sky; at low latitudes we expect the systematics in the map to be somewhat worse, as the Galactic and stellar models we use were trained at high latitudes \citep{Green:2014}.

\begin{figure}[htb]
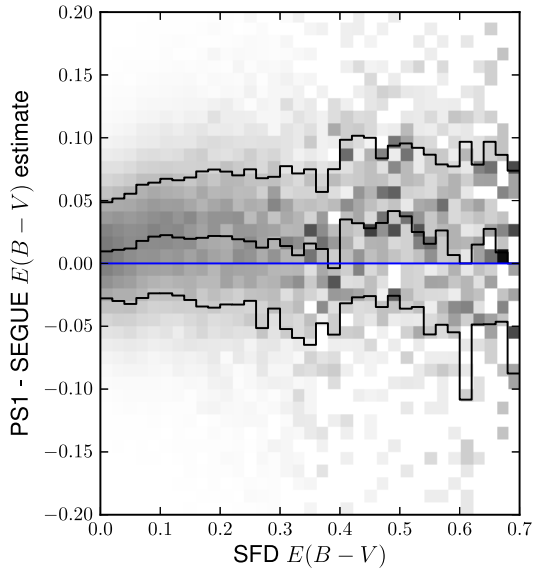

\dfplot{segue}
\figcaption{
\label{fig:segue} PS1 reddening map compared with reddening estimates toward stars observed spectroscopically by the SDSS having $|b| > 20\degree$.  The distribution of differences between the two reddening estimates are shown as a function of the SFD reddening toward each star.  Apart from a $\sim20$~mmag overall offset, the agreement is excellent and close to the uncertainty in the SEGUE reddenings.
}
\end{figure}

Ideally, we would like to assess our accuracy by means of a comparison with an external set of independent reddening measurements.  The work of \citet{Schlafly:2011} provides color excess measurements for hundreds of thousands of stars observed as part of the SEGUE survey \citep{Yanny:2009}.  We show in Figure~\ref{fig:segue} the comparison between our reddening map and the reddening estimates to the SEGUE stars, using their $g-r$ colors; the result is insensitive to the choice of color.  We here use stars only with $|b| > 20\degree$, to ensure that the stars are behind all the dust.  The Figure shows the distribution of differences between the SEGUE and PS1 reddening estimates, binned by their SFD-estimated reddening.  Solid lines show the 16th, 50th, and 84th percentiles of the distribution.  Perfect agreement is indicated by the blue horizontal line.  

The agreement between our map and the SEGUE reddenings is excellent.  There is a small 10--20 mmag overall offset between the measurements, which is within the systematic uncertainties of our method.  The scatter between the estimates is only 42 mmag.  Since the SEGUE estimates themselves have a typical uncertainty of 33 mmag, the implied uncertainty in our map is only about 25 mmag, comparable to our formal uncertainty estimates.  We note that this test is similar to that in \citet{Green:2014} (\textsection~7.3 of that work), except that test compares the SEGUE data with reddening estimates for individual stars, while this test compares the SEGUE data with the map derived from hundreds of individual stars in each pixel.  

This accuracy is the best of which we are aware in large scale reddening maps.  For comparison, the 2MASS-based map of \citet{Lombardi:2011} near the Galactic anticenter has $3^\prime$ FWHM and $\sigma_{E(B-V)}$ of 90 mmag.  Scaling the noise in our map up by a factor of two to account for the worse resolution of our map (half the resolution in the $|b| < 30\degree$ sky), our map still has significantly better signal to noise than the 2MASS map.  This reddening sensitivity is a consequence of the deeper PS1 imaging and the use of optical colors.  However, because our map relies on optical data, it saturates at an $E(B-V)$ of about 1.5 mags, while the near-infrared map of \citet{Lombardi:2011} can reach $E(B-V) > 15$.

\section{Discussion}
\label{sec:discussion}

Our map directly measures the reddening from dust over three-quarters of the sky.  We compare our map with the FIR-emission-based SFD map and the more recent {\it Planck} dust map \citep{Planck:2013}, to get a sense for the trade-offs when estimating extinction directly or indirectly through the dust emission.  Emission-based maps provide much higher signal-to-noise, but systematic errors in the conversion from emission to reddening often reduce this benefit.

We note three challenges in the comparison of emission-based maps and the PS1 map.  First, our analysis assumes that the reddening is always positive.  This biases our reddening estimates in the low $E(B-V)$ sky high by about 20~mmag.  Second, the PS1 map tracks dust reddening at 4.5~kpc.  Meanwhile, the emission-based maps track the total column in each pixel.  At low latitudes, there may be substantial dust beyond 4.5~kpc, present in emission-based maps but not in the PS1 map.  At high latitudes, however, nearly all of the dust should be well within 4.5~kpc, allowing comparison between the maps.  Third, our map tracks the typical reddening of stars observed behind the dust, while emission-based maps track the PSF-smoothed total column density.  In filamentary regions that contain significant structure on scales smaller than one pixel ($7^\prime$--$14^\prime$ in the PS1 map), the reddening of the typical star can be significantly different from the PSF-smoothed column in a pixel.

\subsection{Comparison with SFD}
\label{subsec:sfd}

\begin{figure*}[htb]
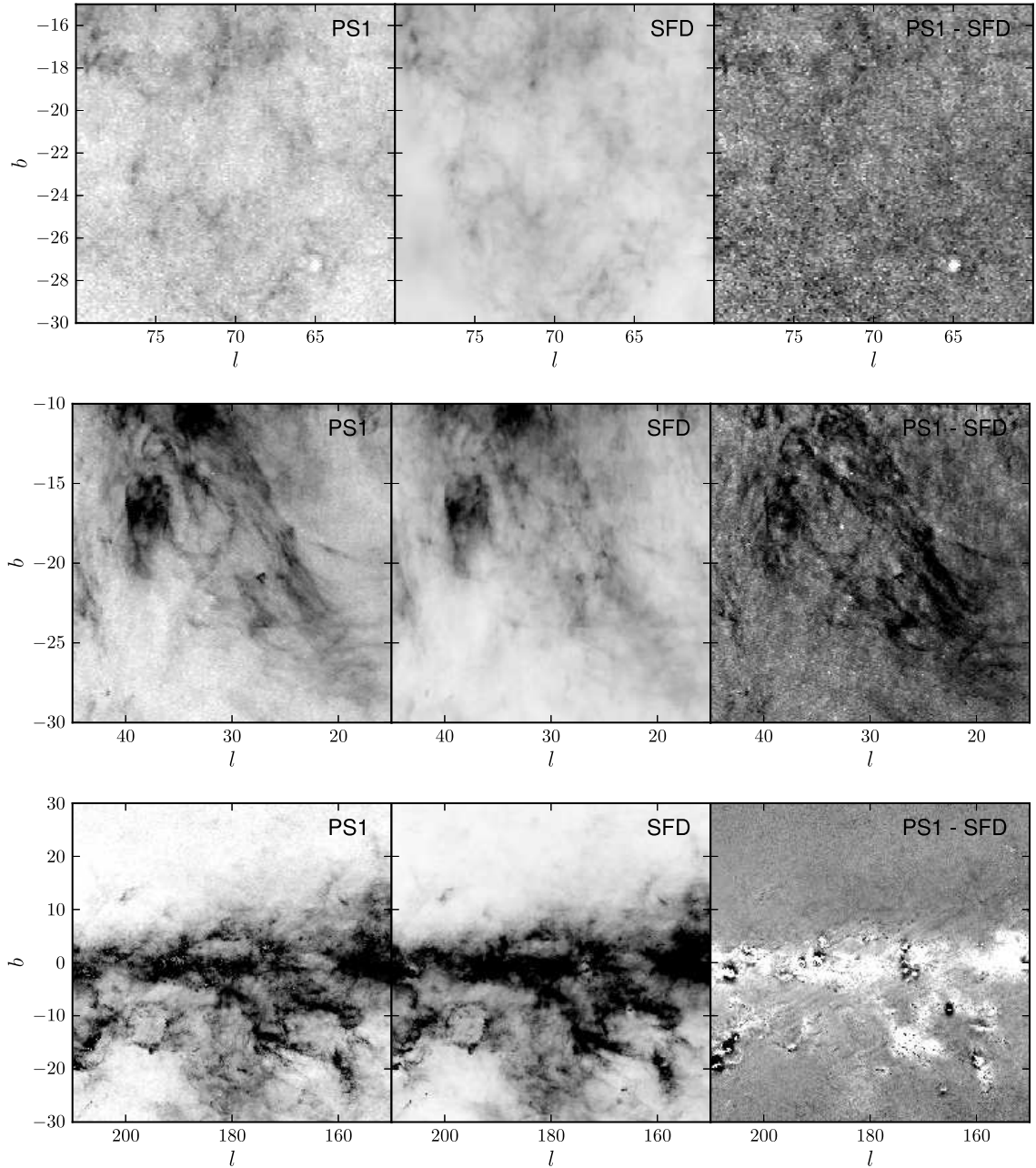

\dfplot{ext-sfd-zoomsouth2}
\dfplot{ext-sfd-zoomsouth}
\dfplot{ext-sfd-zoomac}
\figcaption{
\label{fig:zoompanel}  PS1 reddening map (left), SFD dust map (middle), and the difference (right), for three regions of interest: a relatively blank field in the south Galactic cap (top), a region around the Aquila South cloud (middle), and a region in the Galactic anticenter (bottom).  The grayscale in the first two rows ranges from 0--0.5 mags $E(B-V)$ for the maps and $-0.1$--$0.1$ mags for the differences, while the last row ranges from 0--1 mag $E(B-V)$ for the maps and $-0.25$--$0.25$ mags for the differences.  We have removed a 30 mmag overall offset from the PS1 maps to highlight the structure in the maps, rather than overall offsets.
}
\end{figure*}

Figure~\ref{fig:zoompanel} compares the PS1 reddening map with the SFD reddening map in a few regions of interest.  Qualitatively, the two maps closely agree in all cases.  At high Galactic latitudes, where the dust is well within 4.5~kpc, we expect the best agreement between the maps.  This good agreement is verified by the first row of panels in Figure~\ref{fig:zoompanel}.  For $60\degree < l < 80\degree$, $-30\degree < b < -15\degree$, the two maps have an rms difference of only 25 mmag---a value which includes uncertainty in both maps.  In the high-latitude $E(B-V) < 0.2$ sky, agreement of 20--30 mmag is typical.

The second row of panels of Figure~\ref{fig:zoompanel} shows another high Galactic latitude region, this time highlighting a more interesting part of the sky: the Aquila South cloud \citep{Dame:2001}.  Here again, PS1 and SFD are qualitatively in close agreement.  However, the difference map makes clear that SFD is missing a large, filamentary structure that is present in the PS1 map.  The reddening in this cloud, absent in SFD, can be quite large: as much as about 0.1 mag $E(B-V)$.  A likely explanation for the discrepancy is that the dust in these filaments is substantially colder than the surrounding dust.  The reddening $E(B-V)$ was modelled by SFD as 
\begin{equation}
\label{eq:sfdtemp}
E(B-V) = 0.0184~X_{\mathrm{SFD}}~I_{100}
\end{equation}
where $I_{100}$ is the zodiacal-light-subtracted composite IRAS-DIRBE 100\mum\ map, and $X_\mathrm{SFD}$ is a temperature-correction factor derived from the DIRBE 100\mum\ and 240\mum\ maps, normalized to be one at a temperature typical of high-latitude dust.  The DIRBE maps have an angular resolution of only about 1\degree, so these filaments are not resolved in the SFD temperature map.  Accordingly, SFD assigns the filaments temperatures typical of the surrounding warmer dust, leading to underestimated reddening.  This conclusion is borne out by comparison with the {\it Planck} maps in the region, which agree better with the PS1 map than SFD does.

The third row of panels shows the maps in the direction of the Galactic anticenter.  While the maps agree qualitatively, large residuals are present in the difference map.  Dust beyond 4.5~kpc contributes to these residuals, though other residuals are also prominent.  The largest differences take the form of a bright point with a dark halo.  This is a signature of bright FIR point sources lighting up a few pixels in the SFD dust map, and then artificially boosting the inferred temperature of the surrounding dust, leading to depressed reddening estimates in the vicinity.  This effect is visible directly in SFD alone around $(l, b) = (165\degree, -9\degree)$, where the FIR-bright HII region NGC 1579 has drilled a small region out of the map.

\begin{figure*}[htb]
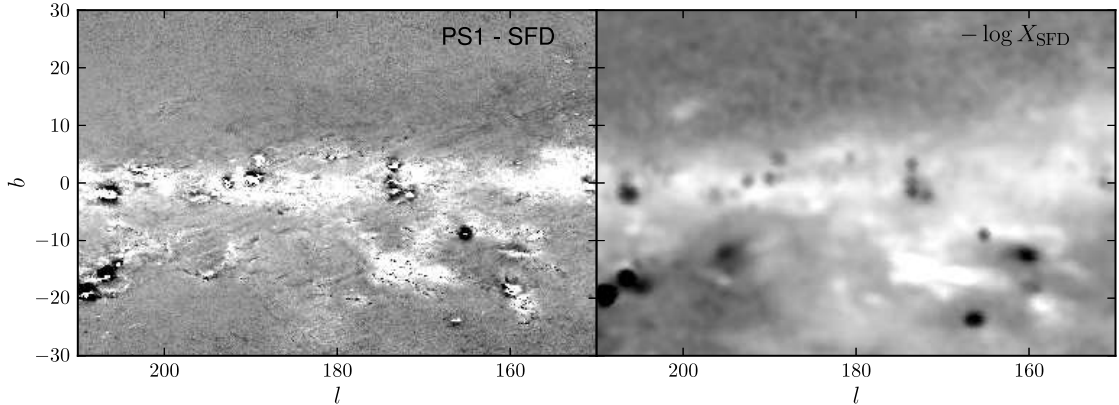

\dfplot{ext-sfd-zoomactemp}
\figcaption{
\label{fig:tempac}  Difference between PS1 and SFD reddening maps (left), compared with the SFD temperature correction factor (right).  The strong correlation between the maps shows that the SFD temperature correction factor is problematic.
}
\end{figure*}

Another salient feature of the maps of the Galactic anticenter is that the difference map strikingly resembles the SFD temperature-correction map.  Figure~\ref{fig:tempac} shows the difference map toward the Galactic anticenter and the $X_\mathrm{SFD}$ temperature-correction factor (Equation \ref{eq:sfdtemp}).  The strong correlation indicates that $X_\mathrm{SFD}$ could be dramatically improved, as has earlier been suggested \citep{Schlafly:2010, Peek:2010}.  Much of the challenge appears to stem from the low resolution of $X_\mathrm{SFD}$.  This is particularly exciting given the recent or forthcoming availability of higher-resolution WISE, AKARI, and {\it Planck} maps of dust emission; the combination of these data represent a tremendous advance over earlier IRAS and DIRBE data.  Indeed, comparison with {\it Planck} (\textsection\ref{subsec:planck}) indicates that substantial progress has been made.

\subsection{Comparison with {\it Planck}}
\label{subsec:planck}

The recent \citet{Planck:2013} dust map combines far-infrared measurements from \citet{Schlegel:1998} with {\it Planck} microwave data to more completely trace the spectral energy distribution of the Galaxy's dust.  The improved wavelength coverage and resolution of the {\it Planck} data relative to the {\it COBE/DIRBE} data used by \citet{Schlegel:1998} for studying dust temperature should allow for dramatic improvements in dust mapping.

The \citet{Planck:2013} provide two estimates of dust column density: one based on the total dust radiance, $\mathcal{R}$, and another based on the optical depth of the dust at 353~GHz, $\tau_{353}$.  We have adopted the calibration of \citet{Planck:2013} to convert each of these maps into $E(B-V)$ for comparison.  Throughout, when referring to the $\mathcal{R}$ or $\tau_{353}$ maps, we are referring to these maps after calibration into units of $E(B-V)$.  The dust radiance $\mathcal{R}$ is defined as $\mathcal{R} = \int I_\nu d\nu$, the integral of the dust emission spectrum over frequency.  This quantity is not obviously appropriate as a reddening template, given its dependence on the interstellar radiation field (ISRF) and the corresponding temperature of the dust.  The primary motivations for adopting $\mathcal{R}$ as a probe of the dust column instead of $\tau_{353}$ are that it is less sensitive to degeneracies in the fit parameters used in the {\it Planck} dust fits, and that it is less affected by the cosmic infrared background anisotropy.  Due to the ISRF dependence of the $\mathcal{R}$ map, \citet{Planck:2013} recommends using $\tau_{353}$ instead of $\mathcal{R}$ to trace dust column in translucent and dense clouds where the ISRF is significantly modified.

The accuracy of the \citet{Planck:2013} dust maps is largely tested by comparison to $N_H$, as $N_H$ is the only all-sky map which is a good proxy for dust column.  Calibration of the {\it Planck} map to reddening $E(B-V)$ was made using a sample of quasars probing the very low extinction sky ($E(B-V) \lesssim 0.1$).  Our current reddening map covers three-quarters of the sky and $E(B-V)$ up to about 1.5~mag before saturating, allowing a much broader test.

\subsection{The Overall $E(B-V)$ Scale}
\label{subsec:ebvscale}
A basic test is verifying that the various maps share the same $E(B-V)$ scaling.  The PS1 map measures reddening in the PS1 bands, rather than $E(B-V)$ directly.  It is tied to $E(B-V)_\mathrm{SFD}$ according to the reddening vector of \citet{Schlafly:2011}.  Provided that \citet{Schlafly:2011} is correct, this places our map and SFD on the same overall scale.  However, \citet{Schlafly:2011} finds that the $E(B-V)$ scale of SFD is off by 14\%.  Our map inherits that offset, though see below for further discussion.

\begin{figure}[htb]
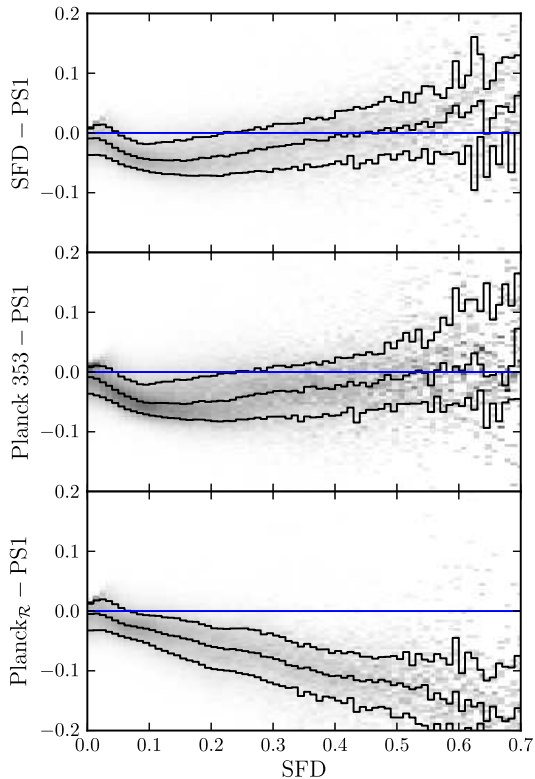

\dfplot{ebvcomp}
\figcaption{
\label{fig:ebvcomp}  The distribution of differences between SFD, {\it Planck} $\tau_{353}$, and {\it Planck} $\mathcal{R}$ as a function of SFD, for the high-latitude sky ($|b| > 30\degree$).  Systematic differences between the PS1 map and emission-based maps exist for all maps.
}
\end{figure}

We show the distribution of differences between our map, SFD, and the {\it Planck} $\tau_{353}$ and $\mathcal{R}$ maps as a function of SFD in Figure~\ref{fig:ebvcomp}.  As in Figure~\ref{fig:segue}, the solid black lines indicate the 16th, 50th, and 84th percentiles of the distribution in each bin, and the solid blue line indicates perfect agreement.  We include here only the $b > 30\degree$ sky to ensure that essentially all of the dust is within 4.5~kpc, and to focus on the region of sky where the agreement between the PS1 map and the emission-based maps is best.  Figure~\ref{fig:ebvcomp} shows that the PS1 map and the emission maps systematically disagree as a function of $E(B-V)$.

There is a clear residual trend in $\mathrm{SFD}-\mathrm{PS1}$ and $\tau_{353}-\mathrm{PS1}$ with the sense that, at reddenings less than $0.15$~mag, PS1 finds more reddening per unit SFD than the emission-based maps, while at larger $E(B-V)$, PS1 finds less reddening per unit SFD.  That is, the slope of the residuals in Figure~\ref{fig:ebvcomp} is negative for $E(B-V) < 0.1$, and positive for $E(B-V) > 0.2$.  This trend is not expected; {\it a priori} we would have expected PS1 to be biased high at low reddening, leading to the opposite trend at low $E(B-V)$.  That effect is however very small, and causes only the diagonal envelope of points with $E(B-V)_\mathrm{SFD} < 0.03$ and positive residuals in each panel of Figure~\ref{fig:ebvcomp}.  The {\it Planck} $\mathcal{R}$ map meanwhile has a different residual trend, with a constant negative slope with no change in behavior around $E(B-V) = 0.15$ mag.

\begin{figure}[htb]
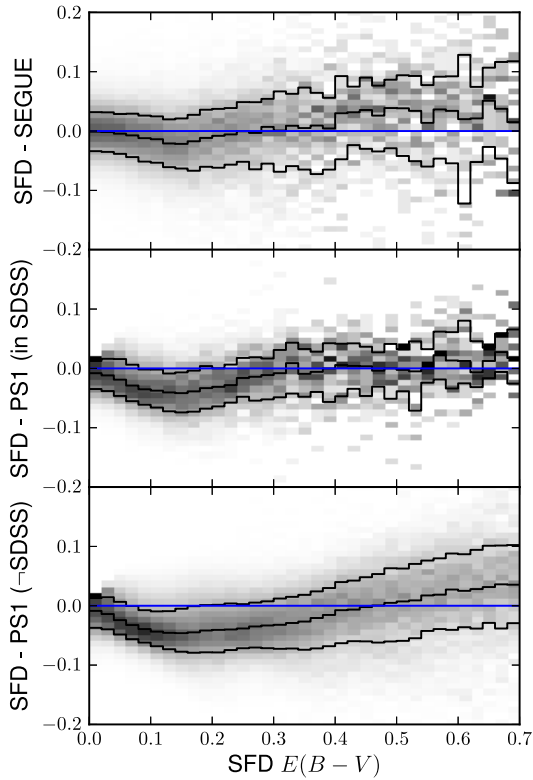

\dfplot{seguemore}
\figcaption{
\label{fig:seguemore} Comparison between the SEGUE and SFD reddening measurements on the $|b| > 20\degree$ sky sampled by SEGUE (top panel); between PS1 and SFD over the same sky footprint (middle panel); and between SFD and PS1 over the $|b| > 20\degree$ sky outside the SEGUE footprint (bottom panel).  While there is good agreement in overall scale between SFD and PS1 within the SDSS footprint, outside this region, at $E(B-V)_{\mathrm{SFD}} > 0.3$, the PS1 and SFD reddening estimates have systematically different slopes.
}
\end{figure}

The fact that PS1 finds consistently less reddening than SFD at $E(B-V) > 0.3$ is a surprise.  The PS1 map was intended to inherit the overall reddening scale of \citet{Schlegel:1998}, through its adoption of the reddening vector of \citet{Schlafly:2011}.  Nevertheless, the slope $\Delta\mathrm{SFD}/\Delta\mathrm{PS1}$ and $\Delta\tau_{353}/\Delta\mathrm{PS1}$ is about 1.15 for $E(B-V) > 0.3$.  This result is unexpected given that \citet{Schlafly:2011} find good agreement in scale with SFD, and we find good agreement in scale with \citet{Schlafly:2011} (Figure~\ref{fig:segue}).  The disagreement stems from the different regions of sky used when comparing with SEGUE in Figure~\ref{fig:segue} and between reddening maps in Figure~\ref{fig:ebvcomp}.  The SEGUE comparison is limited to stars observed by SEGUE, within the SDSS footprint, while the reddening map comparison covers essentially the entire $|b| > 30\degree$, $\delta > -30\degree$ sky.  Restricted to the locations on which SEGUE stars were observed, the agreement between SEGUE and SFD is very similar to the agreement between PS1 and SFD.  This is demonstrated in Figure~\ref{fig:seguemore}, which shows the distribution of residuals between SFD, SEGUE, and PS1 over different regions of sky.  The solid black lines give the 16th, 50th, and 84th percentiles of the distribution, while the solid blue line indicates perfect agreement.  The top two panels of Figure~\ref{fig:seguemore} show the differences between SFD and the SEGUE reddening estimates, followed by the differences between SFD and the PS1 reddening estimates, both over the SDSS footprint with $|b| > 20\degree$.  The second panel shows the good agreement in overall scale between SFD and PS1 in this region.  However the third panel shows that in the $|b| > 20\degree$ sky outside the SDSS footprint, there is a clear residual slope between SFD and PS1 at $E(B-V) > 0.3$.  We conclude only that while the calibration of \citet{Schlafly:2011} is appropriate over the SDSS footprint (albeit with 10\% local variations), extrapolation outside of the SDSS footprint is risky as cloud properties can vary substantially over the sky.

The {\it Planck} $\tau_{353}$ map and SFD have very similar residual behavior as a function of $E(B-V)$.  This is expected as both maps are essentially thermal fits to 100\mum\ dust emission and longer wavelength emission.  The {\it Planck} map, however, includes a correction for the variation in the emissivity of the dust, omitted by SFD.  This correction seems to have had only a minor effect on the overall scaling of the {\it Planck} $\tau_{353}$ map when $E(B-V) < 1$.  The cause of the change in residual slope between the PS1 map and both the SFD and $\tau_{353}$ maps around $E(B-V) = 0.15$ is not understood.

The {\it Planck} $\mathcal{R}$ map residuals behave differently.  The map predicts less reddening than the PS1 map everywhere, by a factor of about $1.3$.  Half of this discrepancy may be due to our map's intended tie to $\mathrm{SFD}$ rather than $E(B-V)$.  This leaves a discrepancy of about 15\% which we are unable to explain.  However, it is ultimately a disagreement between the reddening calibration of \citet{Planck:2013} and the reddening measurements of \citet{Schlafly:2011}, rather than a problem in the technique we have adopted to map dust using PS1.  The $\mathcal{R}$ residuals show no change in slope at $E(B-V) = 0.15$, unlike the SFD and $\tau_{353}$ residuals.  This suggests that the far-infrared modeling of the dust may be causing the change in slope in the SFD and $\tau_{353}$ residuals.

\subsection{Map Comparisons}
\label{subsec:pmapcomp}

\begin{figure*}[htb]
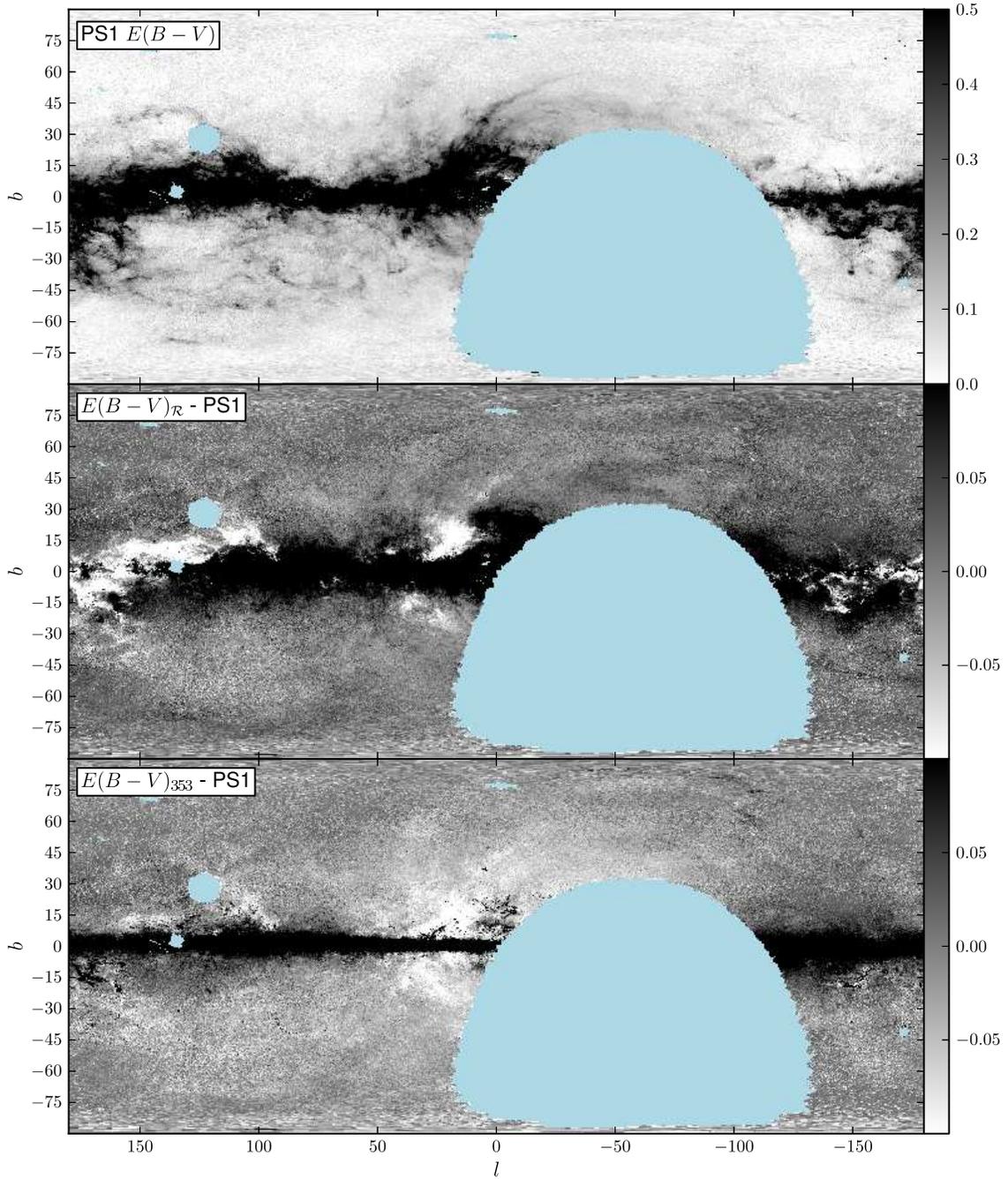

\dfplot{ext-planck-3panelmodscale}
\figcaption{
\label{fig:planckcompare}  Comparison between PS1 and {\it Planck} reddening maps.  The top panel shows our PS1 reddening map, and the following two panels show the difference between it and the {\it Planck} dust-brightness-based and dust-column-based $E(B-V)$ maps.  At high latitudes all maps are qualitatively in good agreement, though in detail we find the {\it Planck} brightness-based map is somewhat more consistent.  Below latitudes of about $20\degree$--$30\degree$, however, the column-based map is much more accurate.
}
\end{figure*}

Figure~\ref{fig:planckcompare} shows our PS1 map, followed by the difference between our PS1 map and the two {\it Planck} reddening maps.  Consistent with \textsection\ref{subsec:ebvscale}, we have rescaled the {\it Planck} $\tau_{353}$ map by 30\% to provide a better match to the PS1 map.  The high-latitude sky is in close agreement, and subtracts well in the difference maps.  However, the lower latitude and higher $E(B-V)$ sky shows large differences between the maps.  Both {\it Planck} maps substantially underpredict the amount of extinction outside the plane east of the the Galactic center, as well as in the Cepheus Flare and at high latitudes towards the anticenter.  Meanwhile at low latitudes the Planck $E(B-V)$ are generally larger than the PS1 $E(B-V)$, which is expected as the PS1 maps saturate at about 1.5~mag and there is often significant dust beyond 4.5~kpc in this area.

In general at low latitudes ($|b| \lesssim 30\degree$), but outside the plane, the PS1 map agrees more closely with the {\it Planck} $\tau_{353}$ map than with $\mathcal{R}$.  This was anticipated by \citet{Planck:2013} given the varying interstellar radiation field in these environments, but our map makes clear that this effect is generic at moderately low latitudes.  Outside of $|b| < 7\degree$ our map and {\it Planck} $\tau_{353}$ agree within about 0.05~mag outside the Galactic center, while the region $7\degree < |b| < 15\degree$ is unreliable in $\mathcal{R}$.  The presence of regions of both overestimated and underestimated reddening shows that even in the high-latitude sky no simple rescaling of the Planck maps can bring them into agreement with the reddening estimates.

\section{Conclusion}
\label{sec:conclusion}

We present a reddening map based on photometry from \PS\ covering almost three-quarters of the sky.  The map has a unique combination of high angular resolution ($7^\prime$--$14^\prime$ pixels), low noise ($\sim 25~\mathrm{mmag}~E(B-V)$), and wide sky coverage that makes it ideal for studying the Galactic dust over a wide range of environments.

Comparison with the widely-used SFD dust map highlights some known shortcomings of that map.  The combination of low-resolution DIRBE data with high-resolution IRAS data leads to peculiar features in filamentary dust clouds, and can lead to SFD either overestimating or underestimating the true reddening.  Comparison with the \citet{Planck:2013} maps shows improvement, but have similar problems despite the wider frequency coverage and better resolution.  The {\it Planck} $\mathcal{R}$ map traces the dust well for $|b| > 15\degree$, but is unreliable outside this region due to variation in the interstellar radiation field.

These techniques can be used in any multiband optical survey.  Application to the SkyMapper survey would allow the dust in the remaining quarter of the sky to be mapped.  The Dark Energy Survey will also map much of the remaining sky, and could efficiently map the $\delta < -30\degree$ Galactic plane ($|b| < 5\degree$).  As foreseen by \citet{BailerJones:2011}, even the photometric component of Gaia alone is extremely interesting from the perspective of this technique---but the addition of parallax information will make that mission truly revolutionary for maps of dust.

ES acknowledges funding by Sonderforschungsbereich SFB 881 ``The Milky Way System'' (subproject A3) of the German Research Foundation (DFG).  DF acknowledges support of NASA grant NNX10AD69G.  GMG and DPF are partially supported by NSF grant AST-1312891.  N.F.M. gratefully acknowledges the CNRS for support through PICS project PICS06183.  Computation was performed on the GPU Cluster Milky Way at FZ Juelich and on the Odyssey cluster supported by the FAS Division of Science, Research Computing Group at Harvard University.

The Pan-STARRS1 Surveys (PS1) have been made possible through contributions of the Institute for Astronomy, the University of Hawaii, the Pan-STARRS Project Office, the Max-Planck Society and its participating institutes, the Max Planck Institute for Astronomy, Heidelberg and the Max Planck Institute for Extraterrestrial Physics, Garching, The Johns Hopkins University, Durham University, the University of Edinburgh, Queen's University Belfast, the Harvard-Smithsonian Center for Astrophysics, the Las Cumbres Observatory Global Telescope Network Incorporated, the National Central University of Taiwan, the Space Telescope Science Institute, the National Aeronautics and Space Administration under Grant No. NNX08AR22G issued through the Planetary Science Division of the NASA Science Mission Directorate, the National Science Foundation under Grant No. AST-1238877, the University of Maryland, and Eotvos Lorand University (ELTE).

%\end{thebibliography}
\bibliography{2dmap}

\begin{thebibliography}{}
\expandafter\ifx\csname natexlab\endcsname\relax\def\natexlab#1{#1}\fi

\bibitem[{{Bailer-Jones}(2011)}]{BailerJones:2011}
{Bailer-Jones}, C.~A.~L. 2011, \mnras, 411, 435

\bibitem[{{Berry} {et~al.}(2012){Berry}, {Ivezi{\'c}}, {Sesar}, {Juri{\'c}},
  {Schlafly}, {Bellovary}, {Finkbeiner}, {Vrbanec}, {Beers}, {Brooks},
  {Schneider}, {Gibson}, {Kimball}, {Jones}, {Yoachim}, {Krughoff}, {Connolly},
  {Loebman}, {Bond}, {Schlegel}, {Dalcanton}, {Yanny}, {Majewski}, {Knapp},
  {Gunn}, {Allyn Smith}, {Fukugita}, {Kent}, {Barentine}, {Krzesinski}, \&
  {Long}}]{Berry:2012}
{Berry}, M., {Ivezi{\'c}}, {\v Z}., {Sesar}, B., {et~al.} 2012, \apj, 757, 166

\bibitem[{{Bressan} {et~al.}(2012){Bressan}, {Marigo}, {Girardi}, {Salasnich},
  {Dal Cero}, {Rubele}, \& {Nanni}}]{Bressan:2012}
{Bressan}, A., {Marigo}, P., {Girardi}, L., {et~al.} 2012, \mnras, 427, 127

\bibitem[{{Burstein} \& {Heiles}(1978)}]{Burstein:1978}
{Burstein}, D., \& {Heiles}, C. 1978, \apj, 225, 40

\bibitem[{{Chabrier}(2001)}]{Chabrier:2001}
{Chabrier}, G. 2001, \apj, 554, 1274

\bibitem[{{Dame} {et~al.}(2001){Dame}, {Hartmann}, \& {Thaddeus}}]{Dame:2001}
{Dame}, T.~M., {Hartmann}, D., \& {Thaddeus}, P. 2001, \apj, 547, 792

\bibitem[{{Dobashi} {et~al.}(2005){Dobashi}, {Uehara}, {Kandori}, {Sakurai},
  {Kaiden}, {Umemoto}, \& {Sato}}]{Dobashi:2005}
{Dobashi}, K., {Uehara}, H., {Kandori}, R., {et~al.} 2005, \pasj, 57, 1

\bibitem[{{Fitzpatrick}(1999)}]{Fitzpatrick:1999}
{Fitzpatrick}, E.~L. 1999, \pasp, 111, 63

\bibitem[{{Gorski} {et~al.}(1999){Gorski}, {Wandelt}, {Hansen}, {Hivon}, \&
  {Banday}}]{Gorski:1999}
{Gorski}, K.~M., {Wandelt}, B.~D., {Hansen}, F.~K., {Hivon}, E., \& {Banday},
  A.~J. 1999, ArXiv Astrophysics e-prints, arXiv:astro-ph/9905275

\bibitem[{{Green} {et~al.}(2014){Green}, {Schlafly}, {Finkbeiner}, {Juri{\'c}},
  {Rix}, {Burgett}, {Chambers}, {Draper}, {Flewelling}, {Kudritzki}, {Magnier},
  {Martin}, {Metcalfe}, {Tonry}, {Wainscoat}, \& {Waters}}]{Green:2014}
{Green}, G.~M., {Schlafly}, E.~F., {Finkbeiner}, D.~P., {et~al.} 2014, \apj,
  783, 114

\bibitem[{{Hanson} \& {Bailer-Jones}(2014)}]{Hanson:2014}
{Hanson}, R.~J., \& {Bailer-Jones}, C.~A.~L. 2014, \mnras, 438, 2938

\bibitem[{{Hodapp} {et~al.}(2004){Hodapp}, {Kaiser}, {Aussel}, {Burgett},
  {Chambers}, {Chun}, {Dombeck}, {Douglas}, {Hafner}, {Heasley}, {Hoblitt},
  {Hude}, {Isani}, {Jedicke}, {Jewitt}, {Laux}, {Luppino}, {Lupton}, {Maberry},
  {Magnier}, {Mannery}, {Monet}, {Morgan}, {Onaka}, {Price}, {Ryan},
  {Siegmund}, {Szapudi}, {Tonry}, {Wainscoat}, \& {Waterson}}]{PS1_optics}
{Hodapp}, K.~W., {Kaiser}, N., {Aussel}, H., {et~al.} 2004, AN, 325, 636

\bibitem[{{Ivezi{\'c}} {et~al.}(2008){Ivezi{\'c}}, {Sesar}, {Juri{\'c}},
  {Bond}, {Dalcanton}, {Rockosi}, {Yanny}, {Newberg}, {Beers}, {Allende
  Prieto}, {Wilhelm}, {Lee}, {Sivarani}, {Norris}, {Bailer-Jones}, {Re
  Fiorentin}, {Schlegel}, {Uomoto}, {Lupton}, {Knapp}, {Gunn}, {Covey},
  {Smith}, {Miknaitis}, {Doi}, {Tanaka}, {Fukugita}, {Kent}, {Finkbeiner},
  {Munn}, {Pier}, {Quinn}, {Hawley}, {Anderson}, {Kiuchi}, {Chen}, {Bushong},
  {Sohi}, {Haggard}, {Kimball}, {Barentine}, {Brewington}, {Harvanek},
  {Kleinman}, {Krzesinski}, {Long}, {Nitta}, {Snedden}, {Lee}, {Harris},
  {Brinkmann}, {Schneider}, \& {York}}]{Ivezic:2008}
{Ivezi{\'c}}, {\v Z}., {Sesar}, B., {Juri{\'c}}, M., {et~al.} 2008, \apj, 684,
  287

\bibitem[{{Juri{\'c}} {et~al.}(2008){Juri{\'c}}, {Ivezi{\'c}}, {Brooks},
  {Lupton}, {Schlegel}, {Finkbeiner}, {Padmanabhan}, {Bond}, {Sesar},
  {Rockosi}, {Knapp}, {Gunn}, {Sumi}, {Schneider}, {Barentine}, {Brewington},
  {Brinkmann}, {Fukugita}, {Harvanek}, {Kleinman}, {Krzesinski}, {Long},
  {Neilsen}, {Nitta}, {Snedden}, \& {York}}]{Juric:2008}
{Juri{\'c}}, M., {Ivezi{\'c}}, {\v Z}., {Brooks}, A., {et~al.} 2008, \apj, 673,
  864

\bibitem[{{Kaiser} {et~al.}(2010){Kaiser}, {Burgett}, {Chambers}, {Denneau},
  {Heasley}, {Jedicke}, {Magnier}, {Morgan}, {Onaka}, \& {Tonry}}]{PS1_system}
{Kaiser}, N., {Burgett}, W., {Chambers}, K., {et~al.} 2010, in Society of
  Photo-Optical Instrumentation Engineers (SPIE) Conference Series, Vol. 7733,
  Society of Photo-Optical Instrumentation Engineers (SPIE) Conference Series,
  77330E

\bibitem[{{Lallement} {et~al.}(2014){Lallement}, {Vergely}, {Valette},
  {Puspitarini}, {Eyer}, \& {Casagrande}}]{Lallement:2014}
{Lallement}, R., {Vergely}, J.-L., {Valette}, B., {et~al.} 2014, \aap, 561, A91

\bibitem[{{Lombardi} \& {Alves}(2001)}]{Lombardi:2001}
{Lombardi}, M., \& {Alves}, J. 2001, \aap, 377, 1023

\bibitem[{{Lombardi} {et~al.}(2011){Lombardi}, {Alves}, \&
  {Lada}}]{Lombardi:2011}
{Lombardi}, M., {Alves}, J., \& {Lada}, C.~J. 2011, \aap, 535, A16

\bibitem[{{Magnier}(2006)}]{PS1_IPP}
{Magnier}, E. 2006, in The Advanced Maui Optical and Space Surveillance
  Technologies Conference, ed. S.~{Ryan} (Kihei, HI: The Maui Economic
  Developer Board), E50

\bibitem[{{Magnier}(2007)}]{PS1_photometry}
{Magnier}, E. 2007, in Astronomical Society of the Pacific Conference Series,
  Vol. 364, The Future of Photometric, Spectrophotometric and Polarimetric
  Standardization, ed. C.~{Sterken}, 153--+

\bibitem[{{Magnier} {et~al.}(2008){Magnier}, {Liu}, {Monet}, \&
  {Chambers}}]{PS1_astrometry}
{Magnier}, E.~A., {Liu}, M., {Monet}, D.~G., \& {Chambers}, K.~C. 2008, in IAU
  Symposium, Vol. 248, IAU Symp. 248, A Giant Step: from Milli- to
  Micro-arcsecond Astrometry, ed. W.~J. {Jin}, I.~{Platais}, \& M.~A.~C.
  {Perryman} (Cambridge: Cambridge Univ. Press), 553--559

\bibitem[{{Majewski} {et~al.}(2011){Majewski}, {Zasowski}, \&
  {Nidever}}]{Majewski:2011}
{Majewski}, S.~R., {Zasowski}, G., \& {Nidever}, D.~L. 2011, \apj, 739, 25

\bibitem[{{Marshall} {et~al.}(2006){Marshall}, {Robin}, {Reyl{\'e}},
  {Schultheis}, \& {Picaud}}]{Marshall:2006}
{Marshall}, D.~J., {Robin}, A.~C., {Reyl{\'e}}, C., {Schultheis}, M., \&
  {Picaud}, S. 2006, \aap, 453, 635

\bibitem[{{Nidever} {et~al.}(2012){Nidever}, {Zasowski}, \&
  {Majewski}}]{Nidever:2012}
{Nidever}, D.~L., {Zasowski}, G., \& {Majewski}, S.~R. 2012, \apjs, 201, 35

\bibitem[{{Onaka} {et~al.}(2008){Onaka}, {Tonry}, {Isani}, {Lee}, {Uyeshiro},
  {Rae}, {Robertson}, \& {Ching}}]{PS1_GPCB}
{Onaka}, P., {Tonry}, J.~L., {Isani}, S., {et~al.} 2008, in Society of
  Photo-Optical Instrumentation Engineers (SPIE) Conference Series, Vol. 7014,
  Society of Photo-Optical Instrumentation Engineers (SPIE) Conference Series,
  70140D

\bibitem[{{Peek} \& {Graves}(2010)}]{Peek:2010}
{Peek}, J.~E.~G., \& {Graves}, G.~J. 2010, \apj, 719, 415

\bibitem[{{Planck Collaboration} {et~al.}(2011){Planck Collaboration}, {Ade},
  {Aghanim}, {Arnaud}, {Ashdown}, {Aumont}, {Baccigalupi}, {Baker}, {Balbi},
  {Banday}, \& et~al.}]{Planck:2011b}
{Planck Collaboration}, {Ade}, P.~A.~R., {Aghanim}, N., {et~al.} 2011, \aap,
  536, A1

\bibitem[{{Planck Collaboration} {et~al.}(2013){Planck Collaboration},
  {Abergel}, {Ade}, {Aghanim}, {Alina}, {Alves}, {Aniano}, {Armitage-Caplan},
  {Arnaud}, {Ashdown}, \& et~al.}]{Planck:2013}
{Planck Collaboration}, {Abergel}, A., {Ade}, P.~A.~R., {et~al.} 2013, ArXiv
  e-prints, arXiv:1312.1300

\bibitem[{{Rowles} \& {Froebrich}(2009)}]{Rowles:2009}
{Rowles}, J., \& {Froebrich}, D. 2009, \mnras, 395, 1640

\bibitem[{{Sale}(2012)}]{Sale:2012}
{Sale}, S.~E. 2012, \mnras, 427, 2119

\bibitem[{{Sale} {et~al.}(2009){Sale}, {Drew}, {Unruh}, {Irwin}, {Knigge},
  {Phillipps}, {Zijlstra}, {G{\"a}nsicke}, {Greimel}, {Groot}, {Mampaso},
  {Morris}, {Napiwotzki}, {Steeghs}, \& {Walton}}]{Sale:2009}
{Sale}, S.~E., {Drew}, J.~E., {Unruh}, Y.~C., {et~al.} 2009, \mnras, 392, 497

\bibitem[{{Schlafly} \& {Finkbeiner}(2011)}]{Schlafly:2011}
{Schlafly}, E.~F., \& {Finkbeiner}, D.~P. 2011, \apj, 737, 103

\bibitem[{{Schlafly} {et~al.}(2010){Schlafly}, {Finkbeiner}, {Schlegel},
  {Juri{\'c}}, {Ivezi{\'c}}, {Gibson}, {Knapp}, \& {Weaver}}]{Schlafly:2010}
{Schlafly}, E.~F., {Finkbeiner}, D.~P., {Schlegel}, D.~J., {et~al.} 2010, \apj,
  725, 1175

\bibitem[{{Schlafly} {et~al.}(2014){Schlafly}, {Green}, {Finkbeiner}, {Rix},
  {Bell}, {Burgett}, \& {Magnier}}]{Schlafly:2014}
{Schlafly}, E.~F., {Green}, G., {Finkbeiner}, D.~P., {et~al.} 2014, \apj, 786,
  29

\bibitem[{{Schlafly} {et~al.}(2012){Schlafly}, {Finkbeiner}, {Juri{\'c}},
  {Magnier}, {Burgett}, {Chambers}, {Grav}, {Hodapp}, {Kaiser}, {Kudritzki},
  {Martin}, {Morgan}, {Price}, {Rix}, {Stubbs}, {Tonry}, \&
  {Wainscoat}}]{Schlafly:2012}
{Schlafly}, E.~F., {Finkbeiner}, D.~P., {Juri{\'c}}, M., {et~al.} 2012, \apj,
  756, 158

\bibitem[{{Schlegel} {et~al.}(1998){Schlegel}, {Finkbeiner}, \&
  {Davis}}]{Schlegel:1998}
{Schlegel}, D.~J., {Finkbeiner}, D.~P., \& {Davis}, M. 1998, \apj, 500, 525

\bibitem[{{Skrutskie} {et~al.}(2006){Skrutskie}, {Cutri}, {Stiening},
  {Weinberg}, {Schneider}, {Carpenter}, {Beichman}, {Capps}, {Chester},
  {Elias}, {Huchra}, {Liebert}, {Lonsdale}, {Monet}, {Price}, {Seitzer},
  {Jarrett}, {Kirkpatrick}, {Gizis}, {Howard}, {Evans}, {Fowler}, {Fullmer},
  {Hurt}, {Light}, {Kopan}, {Marsh}, {McCallon}, {Tam}, {Van Dyk}, \&
  {Wheelock}}]{Skrutskie:2006}
{Skrutskie}, M.~F., {Cutri}, R.~M., {Stiening}, R., {et~al.} 2006, \aj, 131,
  1163

\bibitem[{{Stubbs} {et~al.}(2010){Stubbs}, {Doherty}, {Cramer}, {Narayan},
  {Brown}, {Lykke}, {Woodward}, \& {Tonry}}]{PS_lasercal}
{Stubbs}, C.~W., {Doherty}, P., {Cramer}, C., {et~al.} 2010, \apjs, 191, 376

\bibitem[{{Tonry} \& {Onaka}(2009)}]{PS1_GPCA}
{Tonry}, J., \& {Onaka}, P. 2009, in Advanced Maui Optical and Space
  Surveillance Technologies Conference,, ed. S.~{Ryan} (Kihei, HI: The Maui
  Economic Developer Board), E40

\bibitem[{{Tonry} {et~al.}(2012){Tonry}, {Stubbs}, {Lykke}, {Doherty},
  {Shivvers}, {Burgett}, {Chambers}, {Hodapp}, {Kaiser}, {Kudritzki},
  {Magnier}, {Morgan}, {Price}, \& {Wainscoat}}]{JTphoto}
{Tonry}, J.~L., {Stubbs}, C.~W., {Lykke}, K.~R., {et~al.} 2012, \apj, 750, 99

\bibitem[{{Yanny} {et~al.}(2009){Yanny}, {Rockosi}, {Newberg}, {Knapp},
  {Adelman-McCarthy}, {Alcorn}, {Allam}, {Allende Prieto}, {An}, {Anderson},
  {Anderson}, {Bailer-Jones}, {Bastian}, {Beers}, {Bell}, {Belokurov},
  {Bizyaev}, {Blythe}, {Bochanski}, {Boroski}, {Brinchmann}, {Brinkmann},
  {Brewington}, {Carey}, {Cudworth}, {Evans}, {Evans}, {Gates}, {G{\"a}nsicke},
  {Gillespie}, {Gilmore}, {Gomez-Moran}, {Grebel}, {Greenwell}, {Gunn},
  {Jordan}, {Jordan}, {Harding}, {Harris}, {Hendry}, {Holder}, {Ivans},
  {Ivezi{\v c}}, {Jester}, {Johnson}, {Kent}, {Kleinman}, {Kniazev},
  {Krzesinski}, {Kron}, {Kuropatkin}, {Lebedeva}, {Lee}, {Leger}, {L{\'e}pine},
  {Levine}, {Lin}, {Long}, {Loomis}, {Lupton}, {Malanushenko}, {Malanushenko},
  {Margon}, {Martinez-Delgado}, {McGehee}, {Monet}, {Morrison}, {Munn},
  {Neilsen}, {Nitta}, {Norris}, {Oravetz}, {Owen}, {Padmanabhan}, {Pan},
  {Peterson}, {Pier}, {Platson}, {Fiorentin}, {Richards}, {Rix}, {Schlegel},
  {Schneider}, {Schreiber}, {Schwope}, {Sibley}, {Simmons}, {Snedden}, {Smith},
  {Stark}, {Stauffer}, {Steinmetz}, {Stoughton}, {Subba Rao}, {Szalay},
  {Szkody}, {Thakar}, {Thirupathi}, {Tucker}, {Uomoto}, {Vanden Berk},
  {Vidrih}, {Wadadekar}, {Watters}, {Wilhelm}, {Wyse}, {Yarger}, \&
  {Zucker}}]{Yanny:2009}
{Yanny}, B., {Rockosi}, C., {Newberg}, H.~J., {et~al.} 2009, \aj, 137, 4377

\bibitem[{{Yasuda} {et~al.}(2007){Yasuda}, {Fukugita}, \&
  {Schneider}}]{Yasuda:2007}
{Yasuda}, N., {Fukugita}, M., \& {Schneider}, D.~P. 2007, \aj, 134, 698

\bibitem[{{York} {et~al.}(2000){York}, {Adelman}, {Anderson}, {Anderson},
  {Annis}, {Bahcall}, {Bakken}, {Barkhouser}, {Bastian}, {Berman}, {Boroski},
  {Bracker}, {Briegel}, {Briggs}, {Brinkmann}, {Brunner}, {Burles}, {Carey},
  {Carr}, {Castander}, {Chen}, {Colestock}, {Connolly}, {Crocker}, {Csabai},
  {Czarapata}, {Davis}, {Doi}, {Dombeck}, {Eisenstein}, {Ellman}, {Elms},
  {Evans}, {Fan}, {Federwitz}, {Fiscelli}, {Friedman}, {Frieman}, {Fukugita},
  {Gillespie}, {Gunn}, {Gurbani}, {de Haas}, {Haldeman}, {Harris}, {Hayes},
  {Heckman}, {Hennessy}, {Hindsley}, {Holm}, {Holmgren}, {Huang}, {Hull},
  {Husby}, {Ichikawa}, {Ichikawa}, {Ivezi{\'c}}, {Kent}, {Kim}, {Kinney},
  {Klaene}, {Kleinman}, {Kleinman}, {Knapp}, {Korienek}, {Kron}, {Kunszt},
  {Lamb}, {Lee}, {Leger}, {Limmongkol}, {Lindenmeyer}, {Long}, {Loomis},
  {Loveday}, {Lucinio}, {Lupton}, {MacKinnon}, {Mannery}, {Mantsch}, {Margon},
  {McGehee}, {McKay}, {Meiksin}, {Merelli}, {Monet}, {Munn}, {Narayanan},
  {Nash}, {Neilsen}, {Neswold}, {Newberg}, {Nichol}, {Nicinski}, {Nonino},
  {Okada}, {Okamura}, {Ostriker}, {Owen}, {Pauls}, {Peoples}, {Peterson},
  {Petravick}, {Pier}, {Pope}, {Pordes}, {Prosapio}, {Rechenmacher}, {Quinn},
  {Richards}, {Richmond}, {Rivetta}, {Rockosi}, {Ruthmansdorfer}, {Sandford},
  {Schlegel}, {Schneider}, {Sekiguchi}, {Sergey}, {Shimasaku}, {Siegmund},
  {Smee}, {Smith}, {Snedden}, {Stone}, {Stoughton}, {Strauss}, {Stubbs},
  {SubbaRao}, {Szalay}, {Szapudi}, {Szokoly}, {Thakar}, {Tremonti}, {Tucker},
  {Uomoto}, {Vanden Berk}, {Vogeley}, {Waddell}, {Wang}, {Watanabe},
  {Weinberg}, {Yanny}, \& {Yasuda}}]{York:2000}
{York}, D.~G., {Adelman}, J., {Anderson}, Jr., J.~E., {et~al.} 2000, \aj, 120,
  1579

\bibitem[{{Yuan} {et~al.}(2013){Yuan}, {Liu}, \& {Xiang}}]{Yuan:2013}
{Yuan}, H.~B., {Liu}, X.~W., \& {Xiang}, M.~S. 2013, \mnras, 430, 2188

\end{thebibliography}

\end{document}